\documentclass[twocolumn,prd,nofootinbib,superscriptaddress]{revtex4}
\usepackage{hyperref}

\newcommand\ForInternalReference[1]{}
\newcommand\SkipForEarlyCirculation[1]{}

\newcommand\SkipPP[1]{}
\usepackage{verbatim}
\usepackage{graphicx}
\usepackage{dcolumn}
\usepackage{bm}
\usepackage{color}
\usepackage{xspace}
\usepackage{url}
\usepackage{amsmath}
\usepackage{float}
\usepackage{multirow}
\usepackage{xcolor}
\usepackage{times}
\newcommand\optional[1]{}

\newcommand\qmstateproduct[2]{\left\langle#1|#2\right\rangle}

\newcommand\unit[1]{{\rm #1}}

\usepackage{color}
\definecolor{amber}{rgb}{1.0, 0.75, 0.0}
\definecolor{orange}{rgb}{1.0, 0.5, 0.0}
\definecolor{amaranth}{rgb}{0.9, 0.17, 0.31}

\graphicspath{{./figures/}}
\newcommand{\mc}{{\cal M}}

\def\ltsima{$\; \buildrel < \over \sim \;$}
\def\simlt{\lower.5ex\hbox{\ltsima}}
\def\gtsima{$\; \buildrel > \over \sim \;$}
\def\simgt{\lower.5ex\hbox{\gtsima}}

\newcommand{\IMRPD}{\textsc{IMRPhenomD}\xspace}

\newcommand{\IMRPDTv}{\textsc{IMRPhenomD\_NRTidalv2}\xspace}

\newcommand{\Resum}{TEOBResumS\xspace}
\newcommand{\ResumS}{TEOBResumS\xspace}
\newcommand{\NRSurTidal}{NRHybSur3dq8Tidal\xspace}
\newcommand\RIFT{RIFT}

\newcommand{\RIT}{\affiliation{Center for Computational Relativity and Gravitation, Rochester Institute of Technology, Rochester, New York 14623, USA}}
\newcommand{\UTA}{\affiliation{Center for Gravitational Physics, University of Texas at Austin, Austin, Texas 78712, USA}}
\begin{document}

\title{Waveform systematics in gravitational-wave inference of signals from binary neutron star merger models incorporating higher order modes information}
%\title{Effects of higher order modes in gravitational-wave inference of signals from binary neutron star mergers}
\author{A.~B.~Yelikar}
\RIT
\author{R.~O'Shaughnessy}
\RIT
\author{J.~Lange}
\UTA
\RIT
\author{A.~Z.~Jan}
\UTA
\RIT

%% PLAN
%% * (a) model averaging at a lower cost
%% * (b) predict biases
%% * (c) propagation into population error, which we show is substantial  (for SPIN)

\begin{abstract}

Accurate information from gravitational wave signals from coalescing binary neutron stars provides essential input to downstream
interpretations, including inference of the neutron star population and equation of state. 
However, even adopting the currently most accurate and physically motivated models available for parameter estimation
(PE) of BNSs, these models remain subject to waveform modeling uncertainty: differences between these models may
introduce biases in recovered source properties.  In this work, we describe injection studies investigating these systematic differences between the two best waveform models available for BNS currently, \NRSurTidal and \ResumS. We demonstrate that for BNS sources observable by current second-generation detectors, differences for low-amplitude signals are significant for certain sources. 

%% This project aims to assess the nature and impact of systematic differences between different 
%% waveform approximations for a binary black hole, binary neutron star, and black hole/neutron star mergers.
%% \textcolor{red}{More content needed!!}
\end{abstract}
\maketitle

\section{Introduction}
\label{sec:Intro}

% Boilerplate
%  - GW observations
%  - systematic error: exemplified with differences seen for 190521g, 190412m, and forthcoming GWTC-1
%  - waveform accuracy studies show it will still matter in the near future (previous work) ...
%\editremark{Blue- copied from BBH marg paper }

Since the discovery of gravitational waves from  GW150914 \cite{DiscoveryPaper}, the Advanced Laser Interferometer Gravitational-Wave Observatory (LIGO) 
~\cite{2015CQGra..32g4001L} and  Virgo
\cite{gw-detectors-Virgo-original-preferred,TheVirgo:2014hva}
detectors continue to discover gravitational waves (GW) from coalescing binary black holes (BBHs) and neutron stars.
The properties of each source are inferred by comparing each observation to some estimate(s), commonly called an approximant, 
for the GWs emitted when a BBH merges.  
As illustrated recently with GW190521 \cite{LIGO-O3-GW190521-discovery,LIGO-O3-GW190521-implications},
GW190814 \cite{LIGO-O3-GW190814}, GW190412 \cite{LIGO-O3-GW190412}, and the discussion in GWTC-3 \cite{LIGOScientific:2021djp}, these
approximations have enough differences with respect to to each other to produce noticeable differences in inferred posterior distributions, consistent with prior work 
\cite{Shaik:2019dym,gwastro-Systematics-Williamson2017,Purrer:2019jcp}.
Despite ongoing generation of new waveforms with increased accuracy \cite{gwastro-mergers-IMRPhenomP,gwastro-mergers-IMRPhenomPv3,gwastro-SEOBNRv4,2019arXiv190509300V,gwastro-mergers-IMRPhenomXP,2020PhRvD.102d4055O}, these previous investigations suggest that waveform model systematics
can remain a limiting factor in inferences about individual events \cite{Shaik:2019dym} and populations 
\cite{gwastro-PopulationReconstruct-Parametric-Wysocki2018,Purrer:2019jcp}.

Waveform systematics could be particularly pernicious for detailed analyses to infer the nuclear equation of
state from GW observations. For analyses not involving postmerger physics, these approaches look for the subtle impact
of matter on the pre-merger inspiral radiation, due to tidal deformations and altered inspiral rate \cite{2021PhRvD.103l4015G,2022arXiv220506023N,2022PhRvD.105h4021C,2022PhRvD.105f1301K,2021arXiv211109214A,2021arXiv210407533R}. 
Even though the GW signal from the early inspiral is well understood because tidal effects are small and accumulate only at the very end of the
inspiral, they're embedded deep within the most challenging strong field component of the GW signal.  
%\editremark{now do corresponding discussion which is BNS focused} CITE BNS models
One known limitation of most previous investigations of waveform systematics for BNS is the neglect of  higher-order
modes (HOM). Current state-of-the-art BNS models \Resum~\cite{2018PhRvD..98j4052N} and \NRSurTidal~\cite{Barkett:2019tus} incorporate higher-order-modes enabling the exploration of these effects.  
%\editremark{Note we will need to talk about \cite{2021arXiv211109214A}, which is like what we want to do -- equivalent
%  to Ashton/Khan, but for us, maybe more work is out from same-Check}. 
%
For example, GW190412~\cite{LIGO-O3-GW190412}, which was a merger of two black holes that were highly asymmetric in 
masses, 30$M_{\odot}$ and 8$M_{\odot}$, demonstrated the existence and importance of HOM in parameter inference of GW from binary mergers~\cite{Islam:2020reh}.  
%The HOM provided more accurate source parameters, including masses, distance, and inclination.
Using models that incorporate HOM can significantly impact the
inferred parameters of sources identified with current-generation instruments for GW170817 or GW170817-like signals, as demonstrated in ~\cite{2022APS..APRD17007L,CalderonBustillo:2020kcg}. Despite their expected significance to parameter inference, most studies of BNS systematics omit them and rarely 
perform large-scale parameter inference studies to fully assess the impact of systematics, although a similar study was done in ~\cite{Huang:2020pba}.
%\editremark{Sylvia and Salvo did something for 3G I think recently -- find this? 2022MNRAS.511.4350B}
For example, several mismatch studies are mostly done
for models having only leading-order (2,2) mode~\cite{Samajdar:2018dcx,Samajdar:2019ulq,Dietrich:2020eud}. 
A study done with fiducial BNS signals with HOMs 
argued that biases in inferring the reduced tidal parameter could be larger than the statistical 90$\%$ only for very high 
SNR signals $\sim$80~\cite{Gamba:2020wgg} in the LIGO-Virgo band. 
Recent work by Narikawa~\cite{Narikawa:2023deu} looked at the effects of multipoles by comparing MultipoleTidal model to PNTidal
and NRTidalv2 waveforms, showing that mismatches and phases do differ between them for systems with higher mass and 
large tidal deformabilities.

This paper is organized as follows.
In Section \ref{sec:methods}, we review the use of RIFT for parameter inference; the waveform models used in this
work; and the techniques used in \cite{Jan:2020bdz} to assess systematic error. 
We describe one fiducial ensembles of synthetic sources,  targeted at the most common  (low) amplitude sources.
In Section \ref{sec:results}, we use two well-studied waveform models to demonstrate the impact of contemporary model
systematics. We show that model systematics will be
important, at a level which must impact population results and consistency tests like PP plots.   
In Section \ref{sec:conclude}, we summarize our results and discuss their potential applications to future GW sources and
population inference.

\section{Methods}
\label{sec:methods}

\subsection{RIFT review}
\label{subsec:RIFT}

A coalescing compact binary in a quasi-circular orbit can be completely characterized by its intrinsic
and extrinsic parameters.  By intrinsic parameters, we refer to the binary's detector-frame masses $m_i$, spins $\chi_{i}$, and any quantities
characterizing matter in the system, $\Lambda_i$. By extrinsic parameters, we refer to the seven numbers needed to 
characterize its spacetime location and orientation; luminosity distance ($d_{L}$), right ascension ($\alpha$), declination ($\delta$), inclination ($\iota$), polarization ($\psi$), coalescence phase ($\phi_{c}$), and time ($t_{c}$).
  We will express masses in solar mass units and
 dimensionless nonprecessing spins in terms of cartesian components aligned with the orbital angular momentum
 $\chi_{i,z}$, as we use waveform models that do not account for precession. We will use \bm{$\lambda$}, $\theta$ to
refer to intrinsic and extrinsic parameters, respectively. 

$$ \bm{\lambda} : (\mathcal{M},q,\chi_{1,z},\chi_{2,z},\Lambda_{1},\Lambda_{2}) $$
$$ \theta : (d_{L},\alpha,\delta,\iota,\psi,\phi_{c},t_{c}) $$

RIFT \cite{gwastro-PENR-RIFT,gwastro-RIFT-Update}
consists of a two-stage iterative process to interpret gravitational wave data $d$ via comparison to
predicted gravitational wave signals $h(\lambda, \theta)$. In one stage, for each  $\lambda_\beta$ from some proposed
``grid'' $\beta=1,2,\ldots N$ of candidate parameters, RIFT computes a marginal likelihood 
\begin{equation}
 {\cal L}_{\rm marg}\equiv\int  {\cal L}(\bm{\lambda} ,\theta )p(\theta )d\theta
\end{equation}
from the likelihood ${\cal L}(\bm{\lambda} ,\theta ) $ of the gravitational wave signal in the multi-detector network,
accounting for detector response; see the RIFT paper for a more detailed specification~\cite{gwastro-PENR-RIFT,gwastro-RIFT-Update}.  
In the second stage,  RIFT performs two tasks. First, it generates an approximation to ${\cal L}(\lambda)$ based on its
accumulated archived knowledge of marginal likelihood evaluations 
$(\lambda_\beta,{\cal L}_\beta)$. This approximation can be generated by Gaussian processes, random forests, or other
suitable approximation techniques. Second, using this approximation, it generates the (detector-frame) posterior distribution
\begin{equation}
\label{eq:post}
p_{\rm post}=\frac{{\cal L}_{\rm marg}(\bm{\lambda} )p(\bm{\lambda})}{\int d\bm{\lambda} {\cal L}_{\rm marg}(\bm{\lambda} ) p(\bm{\lambda} )}.
\end{equation}
where prior $p(\bm{\lambda})$ is prior on intrinsic parameters like mass and spin.    The posterior is produced by
performing a Monte Carlo integral:  the evaluation points and weights in that integral are weighted posterior samples,
which are fairly resampled to generate conventional independent, identically distributed ``posterior samples.''
For further details on RIFT's technical underpinnings and performance,   see
\cite{gwastro-PENR-RIFT,gwastro-RIFT-Update,gwastro-PENR-RIFT-GPU,gwastro-mergers-nr-LangePhD}.

\subsection{Waveform models}

The tidal waveform models used in this study are \IMRPDTv, \NRSurTidal,
and \Resum. NRTidalv2 models~\cite{Dietrich:2019kaq} 
are improved versions of NRTidal~\cite{Dietrich:2017aum} models, which 
are closed-form tidal approximants for binary neutron star coalescence and have been analytically added to selected 
binary black hole GW model to obtain a binary neutron star waveform, either in the time or in the frequency domain. 
The \NRSurTidal~\cite{Barkett:2019tus} tidal model is based on the binary black hole hybrid model NRHybSur3dq8, 
which is constructed via an interpolation of NR waveforms. It includes all modes $\ell \leq 4$, ($5,\pm5$) but not ($4,\pm1$) and (4,0) 
and models tidal effect up to $\Lambda_{1,2} < 5000$. This model combines the accuracy of surrogate waveforms with the efficiency of 
PN models. \Resum~\cite{2018PhRvD..98j4052N}  is another but unique time-domain EOB formalism that includes tidal effects for all modes $\ell \leq 4$, but no $\mathit{m}=0$ and 
models tidal effect up to $\Lambda_{1,2} < 5000$ and for spins up to 0.5.

\subsection{Fiducial synthetic sources and PP tests}
\label{sec:sub:pop}

% $30\unit{Mpc}$ and $80\unit{Mpc}$ for high SNR injections and

%\textcolor{blue}{We will only explore the impact of systematics over a limited fiducial population.}
We consider one universe of 100 synthetic signals for a
3-detector network (HLV), with masses
drawn uniformly in $m_i$ in the region bounded by $\mc/M_\odot \in [1.2,1.4]$, $\eta \in [0.2, 0.25]$ and $\Lambda$ 
for each object uniformly distributed up to 1000. These bounds are expressed in terms of $\mc=(m_1m_2)^{3/5}/(m_1+m_2)^{1/5}$ 
and $\eta=m_1m_2/(m_1+m_2)^2$, and encompass the detector-frame parameters of neutron stars observed till date 
\cite{LIGO-GW170817-bns,Abbott:2020uma,LIGOScientific:2021qlt}. 
The extrinsic parameters are drawn uniformly in sky position and isotropically in Euler angles, with source luminosity
distances drawn proportional to $d_L^2$  between $90\unit{Mpc}$ and $240\unit{Mpc}$ for low SNR injections.
% and  between $30\unit{Mpc}$ and $80\unit{Mpc}$ for high SNR injections. 
All our sources have non-precessing spins, with each component assumed to be uniform between $[-0.05,0.05]$, 
this is due to limitations of the  \NRSurTidal model. 

For complete reproducibility, we use  \NRSurTidal, \Resum and \IMRPDTv, starting the 
signal evolution and likelihood integration at $30\unit{Hz}$, performing all analysis with $4096\unit{Hz}$ time series in Gaussian noise with
known advanced LIGO design PSDs \cite{LIGO-aLIGODesign-Sensitivity-Updated}.  The BNS signal is generated for 
300 seconds but analysis was performed only on 128 seconds of data. 
For each synthetic event and interferometer, we use the same noise realization for all waveform 
approximations. Therefore, the differences between them arise solely due to waveform systematics.
The  \NRSurTidal model is utilized with two settings: a) $\ell=5$ and b) $\ell=2$, which includes only the dominant quadrupole mode. 
\Resum and \IMRPDTv approximants are used with $\ell=4$ and $\ell=2$ settings respectively. 

%~\ref{BNS_NRHybTidal_highSNR},

\begin{figure}[ht!]
\centering

\includegraphics[scale=0.45]{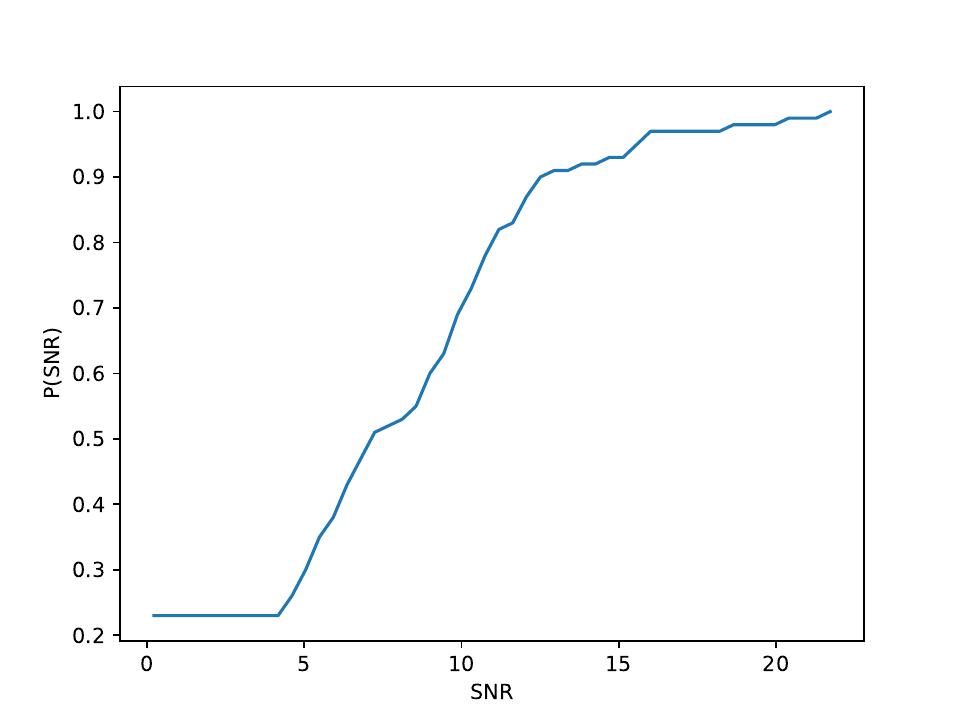}
\caption{Cumulative SNR distribution for a synthetic population of 100 events each drawn from the fiducial BNS populations described in Sec.~\ref{sec:sub:pop}. To avoid ambiguity, this figure shows the expected SNR (i.e., the SNR evaluated using a zero-noise realization).}
\label{BNS_NRHybTidal_combSNR}
\end{figure}

Fig. ~\ref{BNS_NRHybTidal_combSNR} shows the cumulative SNR distribution (under a ``zero-noise" assumption) 
of the specific synthetic population generated from this distribution. Compared to GW170817's confident detection, 
which was a BNS merger that occurred at $40\unit{Mpc}$ detected by LIGO-Virgo with a SNR of 32.4, 
the majority of the signals in this fiducial population have SNRs below or near the typical detection criteria for a BNS merger, 
with some having high enough amplitudes. 

By using a very modest-amplitude population to assess the impact of
waveform systematics, we demonstrate their immediate impact on the kinds of analyses currently being performed on real
observations, let alone future studies.

One way to assess the performance of parameter inference is a probability-probability plots (usually denoted PP-plot) \cite{mm-stats-PP}.
  Using \RIFT{} on each source $k$, with true parameters $\mathbf{\lambda}_k$, we estimate
the fraction of the posterior distributions which is below the true source value $\lambda_{k,\beta}$   [$\hat{P}_{k,\beta}(<\lambda_{k,\beta})$] 
for each intrinsic parameter $\beta$.  After reindexing the sources so $\hat{P}_{k,\beta}(\lambda_{k,\beta})$ increases with $k$ for some
fixed $\beta$, the panels of Figure \ref{pp-samemodel} for example, show a
plot of $k/N$ versus $\hat{P}_k(\lambda_{k,\beta})$ for all binary parameters for different scenarios.

\begin{center}
\begin{table}
\begin{tabular}{ |c|c|c| } 
\hline
Injection model & Recovery model \\
 \hline
 \texttt{NRHybSur3dq8Tidal}($\ell=5$) &  \texttt{NRHybSur3dq8Tidal}($\ell=5$) \\ 
 \texttt{NRHybSur3dq8Tidal}($\ell=5$) & \texttt{NRHybSur3dq8Tidal}($\ell=2$) \\ 
  \texttt{NRHybSur3dq8Tidal}($\ell=5$) & \texttt{TEOBResumS}($\ell=4$) \\ 
 % \texttt{NRHybSur3dq8Tidal}($\ell=5$) & \texttt{TEOBResumS}($\ell=2$) \\ 
  \texttt{NRHybSur3dq8Tidal}($\ell=5$) & \texttt{IMRPhenomD\_NRTidalv2} \\ 
  \texttt{NRHybSur3dq8Tidal}($\ell=5$) & \texttt{IMRPhenomD} \\ 
  \texttt{TEOBResumS}($\ell=4$) & \texttt{TEOBResumS}($\ell=4$) \\
 % \texttt{TEOBResumS}($\ell=4$) & \texttt{TEOBResumS}($\ell=2$) \\
  \texttt{TEOBResumS}($\ell=4$) & \texttt{NRHybSur3dq8Tidal}($\ell=5$) \\
  %\texttt{TEOBResumS}($\ell=4$) & \texttt{NRHybSur3dq8Tidal}($\ell=2$) \\
  \texttt{TEOBResumS}($\ell=4$) & \texttt{IMRPhenomD\_NRTidalv2} \\
   \texttt{TEOBResumS}($\ell=4$) & \texttt{IMRPhenomD} \\
 \hline
\end{tabular}
\caption{List of runs for low-amplitude signals}
\end{table}
\end{center}

\subsection{JS test}
\label{sec:sub:jstest}

To more sharply identify subtle differences introduced by waveform systematics, we will directly compare pairs of inferred posterior probability distributions
deduced with different waveforms but from the same set of data to each other.   Many pairwise error diagnostics have
been used in the literature in general and with RIFT, in particular, \cite{gwastro-RIFT-Update}.   In this study,
motivated by previous work \cite{2021MNRAS.507.2037A}, we use the one-dimensional Jensen-Shannon (JS) divergence
$J(p,q) = (D(q|m) + D(p|m))$ where $D(a|b) = \int dx a(x) \log_2 a(x)/b(x)$  and $m=(a+b)/2$.  The JS divergence is symmetric, ranges between 0 (for
identical distributions) and 1.   For the multidimensional problems described
here, we adopt the median JS divergence over all parameters.
Analyses of O3 using multiple waveforms suggest that binary black holes analyzed with different contemporary waveforms
will produce answers differing by $O(0.02)$  \cite{LIGO-O2-Catalog,Abbott:2020niy,Islam:2023zzj}.

\section{Results}
\label{sec:results}

Our investigations corroborate our central expectation: inferences computed with different waveforms are frequently
substantially different, even for BNS and even for near-threshold events.  The most extreme contrast appeared between
TEOBResumS and other waveform models, where for our near-threshold synthetic events we found ubiquitous qualitatively different
inferences.

\subsection{Anecdotal examples}
\begin{figure}
\includegraphics[scale=0.3]{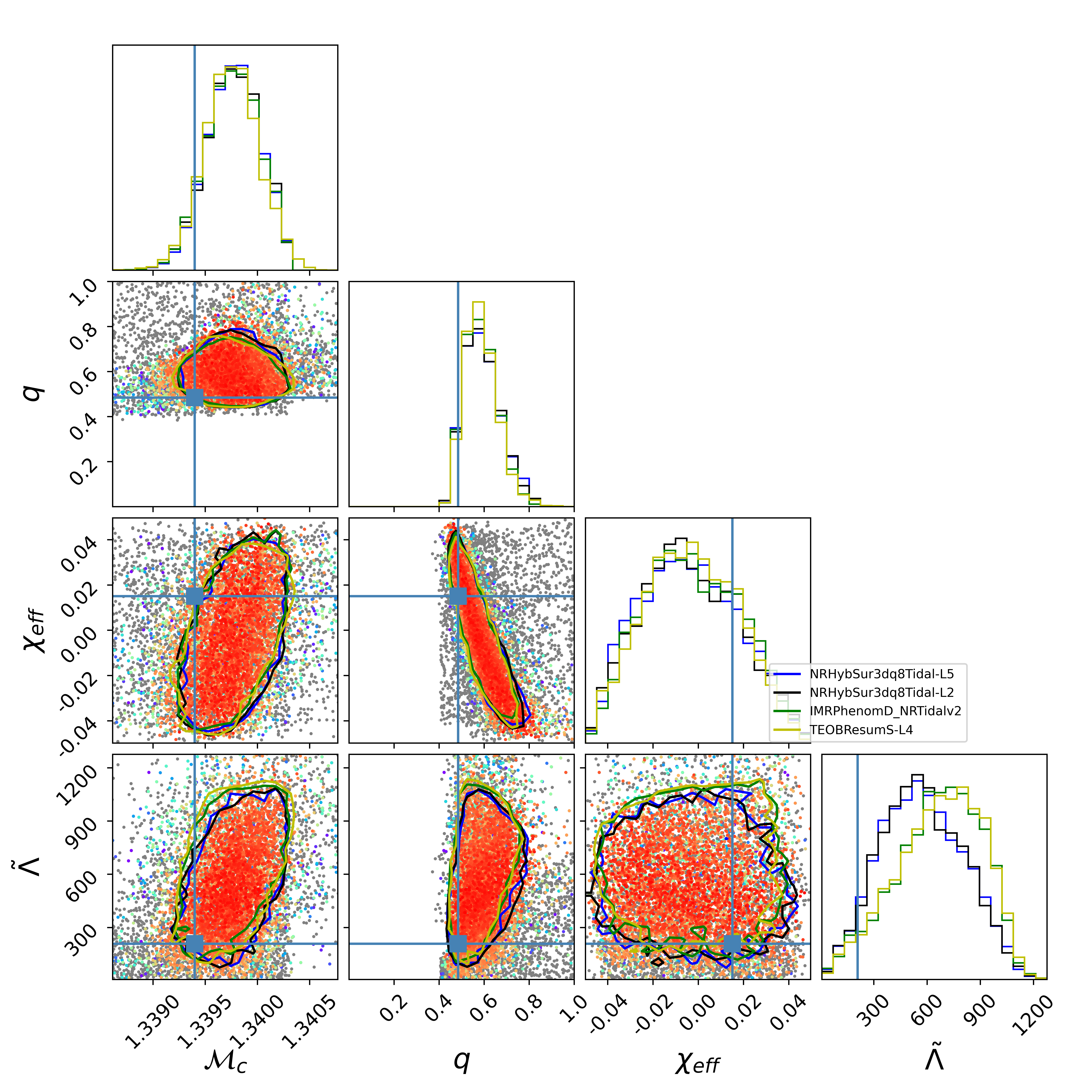}
\caption{Corner plot showing recovered 1D distributions of $\mc$, q, $\chi_{eff}$ and $\widetilde{\Lambda}$ for a low-amplitude NRHybSur3dq8Tidal-lmax5  injection (cross-hairs indicate true value) analysed with various waveform models listed in the legend. The JS-divergence value associated with this event is 0.027.}
\label{corner-sys}
\end{figure}

As an illustrative example of the systematics explored more comprehensively in the population studies below, Figure \ref{corner-sys}
shows the results of parameter inference using multiple recovery waveforms applied to the same synthetic data source,
here a low-amplitude NRHybSur3dq8Tidal-lmax5  injection.   
%FIND A BETTER EXAMPLE
Despite its low signal amplitude, this example shows that posterior distributions derived from the same synthetic data
will differ, depending on the GW signal model used to interpret it.

\subsection{JS divergences: Demonstrating and quantifying waveform systematics}
\label{sec:sub:jsdiv}

\begin{figure}
\includegraphics[scale=0.45]{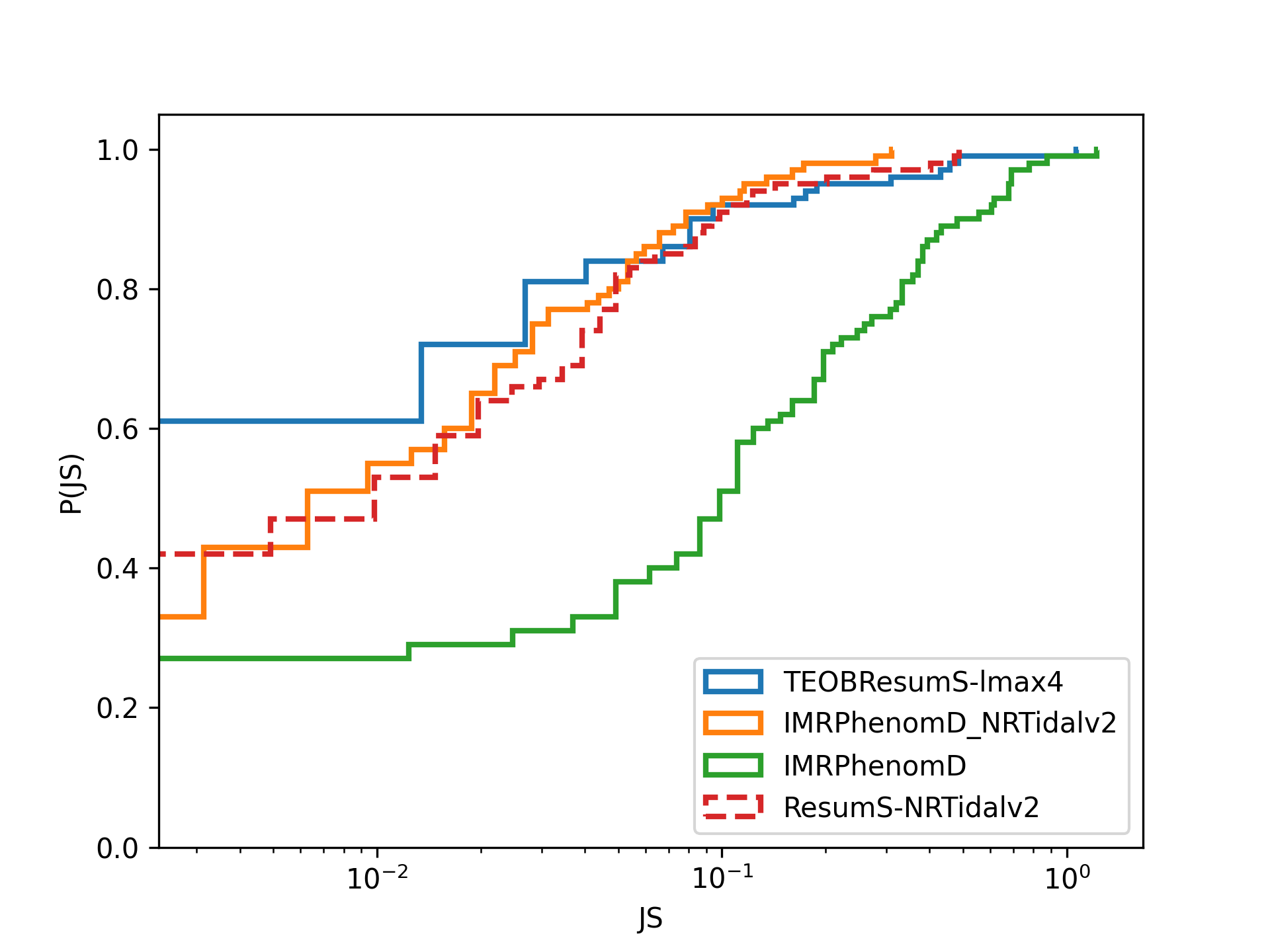}
\includegraphics[scale=0.45]{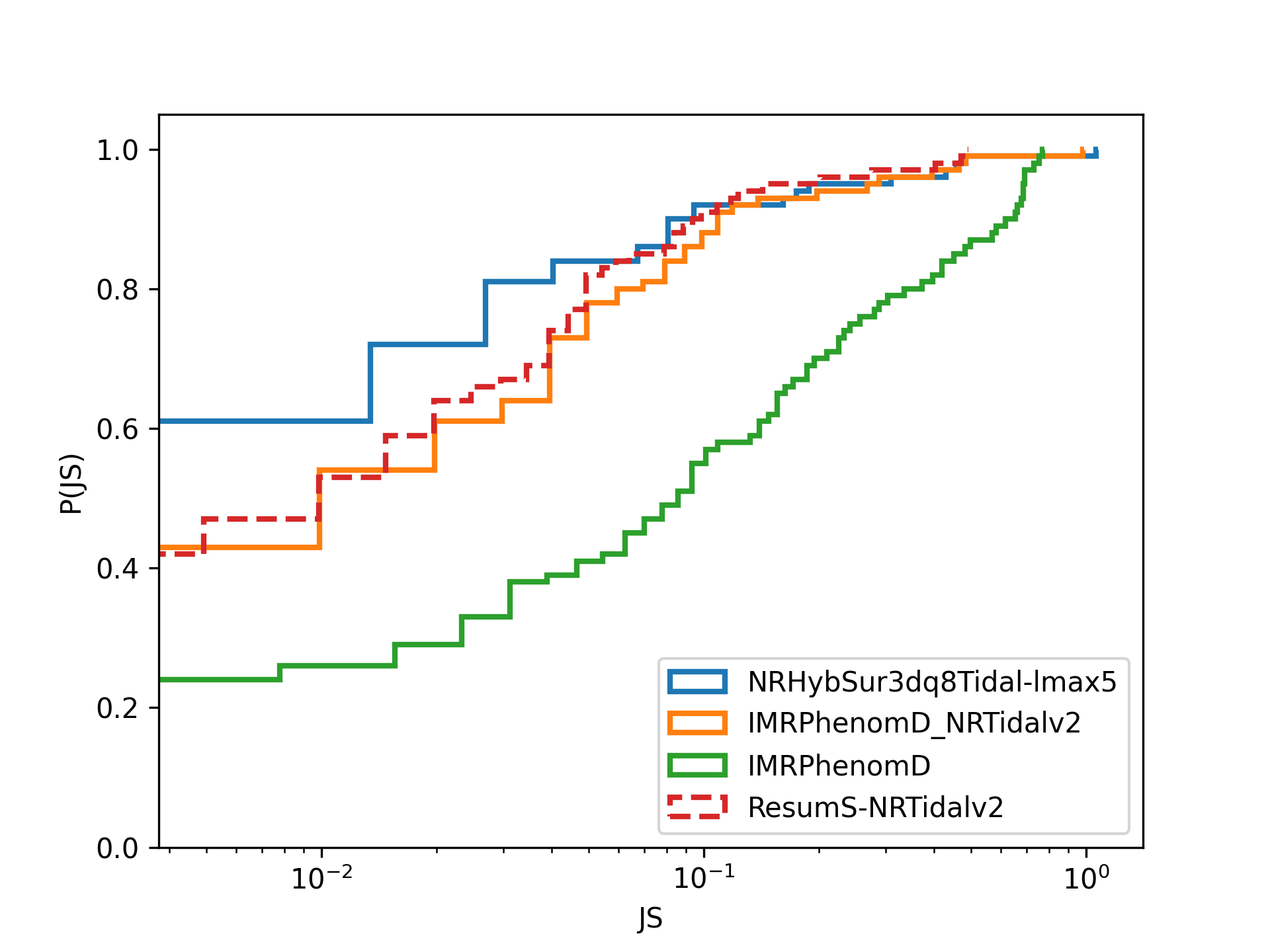}
\caption{JS divergence values of parameters $\mc$, q,  $\chi_{eff}$ and $\widetilde{\Lambda}$ distribution for analysis
  on different recovery waveform models with low-amplitude NRHybSur3dq8Tidal-lmax5 injections.  \emph{Top panel}: Each JS
  divergence used in the CDF is performed between PE constructed  with the indicated model and reference PE constructed
with NRHybSur3dq8.  \emph{Bottom panel}: As above, but using TEOBReumS for reference PE.  }
\label{js-NHSTinj}
\end{figure}

% JS TESTS
Figure \ref{js-NHSTinj} shows the cumulative distribution of combined JS divergences of parameters $\mc$, q, $\chi_{eff}$ 
and $\widetilde{\Lambda}$, between analyses performed with
NRHybSur3dq8Tidal $\ell_{max}=5$ low amplitude injections.   Specifically, the JS divergence is calculated between an
analysis performed using precisely the same model used for injections on the one hand, and the alternative model listed
in the legend on the other.   Inferences performed with all state-of-the-art models that
include tidal physics often produce qualitatively similar inferences, with JS divergences typically less than $10^{-2}$.
Some relatively modest disagreement expected  between (a) different waveform models and (b) the expected modest impact of
higher-order modes for low-mass sources. 
By contrast,  more than 10\% of inferences have JS divergences larger than $10^{-1}$ (mean over all parameters), in all
cases when using models that include similar mode content (but a different waveform model). These calculations suggest that \emph{frequently, both} waveform
systematics \emph{and} higher order modes produce noticably different results.

% (b) include a similar waveform model but different mode content

%
The green line in Figure \ref{js-NHSTinj} shows that tides must be included to avoid significantly biasing the interpretation of a
typical low-amplitude source.  This JS divergence corresponds to inferences that neglect tides entirely (via a
point-particle IMRPhenomD model), even though the true full model includes tides. In this case, the JS divergence is
frequently larger than $0.1$, indicating substantial disagreement with the best possible interpretation.

To investigate the impact of  waveform systematics specifically,  we computed
mismatches between the injected and inference waveforms' (2,2) modes. \emph{Mismatch} is a simple inner-product-based estimate of waveform similarity between two model predictions \cite{gr-nr-WaveformErrorStandards-LBO-2008,2009PhRvD..79l4033R,2010PhRvD..82h4020L,gwastro-mergers-HeeSuk-FisherMatrixWithAmplitudeCorrections,2010PhRvD..82l4052H,2016PhRvD..93j4050K,Purrer:2019jcp}
$h_1(\lambda)$ and $h_2(\lambda)$ at identical model parameters $\lambda$:
\begin{align}
{\cal M}(\lambda) = 1 - \max_{t_c,\phi_c} \frac{|\qmstateproduct{h_1}{e^{i(2\pi f t_c +\phi_c) }h_2}|}{|h_1||h_2|}
\end{align}
In this expression, the inner product
$\langle a|b\rangle_{k}\equiv
\int_{-\infty}^{\infty}2df\tilde{a}(f)^{*}\tilde{b}(f)/S_{h,k}(|f|)$ is  implied by the k$^{th}$ detector's
noise power spectrum $S_{h,k}(f)$, which for the purposes of waveform similarity is assumed to be the
advanced LIGO instrument, H1. In practice, we adopt a low-frequency cutoff $f_{\rm min}$ so all inner products are modified to
\begin{equation}
\label{eq:overlap}
\langle a|b\rangle_{k}\equiv 2 \int_{|f|>f_{\rm min}}df\frac{[\tilde{a}(f)]^{*}\tilde{b}(f)}{S_{h,k}(|f|)}.
\end{equation}

\begin{figure*}

\includegraphics[width=0.45\textwidth]{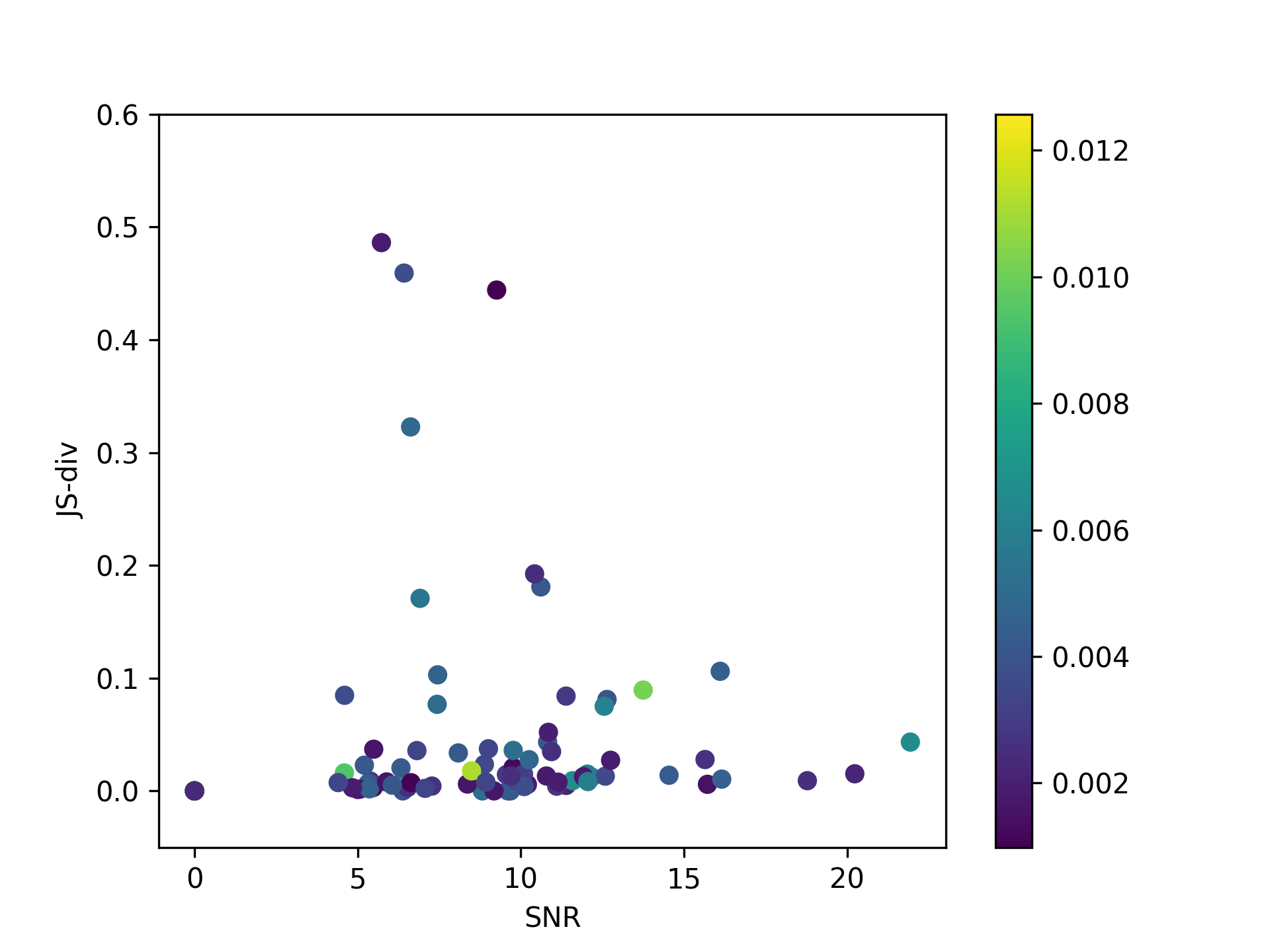}
\includegraphics[width=0.45\textwidth]{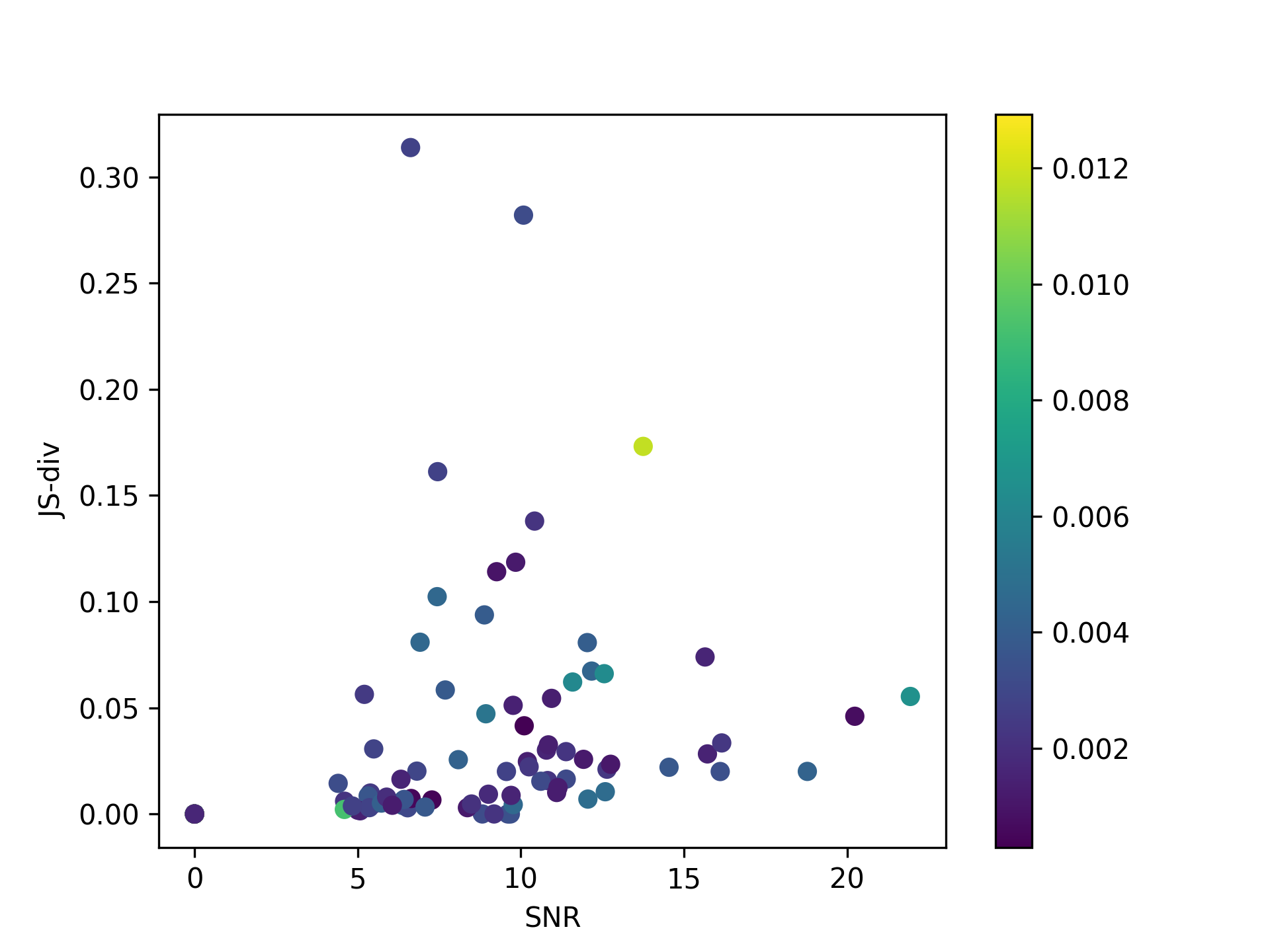}
\caption{\emph{Left panel}: Zero-noise SNR v/s JS divergence (cumulative of $\mc$, q,  $\chi_{eff}$ and
  $\widetilde{\Lambda}$) between \NRSurTidal($\ell_{max}=5$) and \ResumS($\ell_{max}=4$) for the \NRSurTidal
  low-amplitude injections. Color scale shows the mismatch between \NRSurTidal and \Resum.
\emph{Right panel}: As left panel, except for \NRSurTidal and \IMRPDTv. Note the mismatches are calculated for the 
dominant 22-mode only. 
}
\label{fig:js_snr:teob_nrsur}
\end{figure*}

The left panel of Figure
\ref{fig:js_snr:teob_nrsur} shows the results for \NRSurTidal injections recovered with \Resum, with the mismatch shown as a
color scale on top of the injected source SNR and cumulative JS divergence (summed over four one-dimensional JS
divergences for parameters $\mc$, q, $\chi_{eff}$ 
and $\widetilde{\Lambda}$).  While the mismatches are within the waveform accuracy requirements~\cite{2016PhRvD..93d4007K} for most of the injections($>10^{-2}$), higher mismatches don't correlate well with extreme JS divergences.  Rather, below some modest SNR, the random noise realization seems to interact adversely with these large mismatches to produce nearly unconstrained posteriors, such that the similarity
between inferences \emph{becomes stochastic and diverges} at low amplitude. 
The right panel of Figure
\ref{fig:js_snr:teob_nrsur} shows qualitatively similar behavior, using comparisons of \NRSurTidal and \IMRPDTv.

\subsection{PP plots}
The differences between waveforms are significant enough that their imprint can even impact bulk diagnostics such as a PP
plots  \cite{Jan:2020bdz}, which average the impact of waveform systematics over a large population of randomly chosen events.

\begin{figure*}
\includegraphics[scale=0.5]{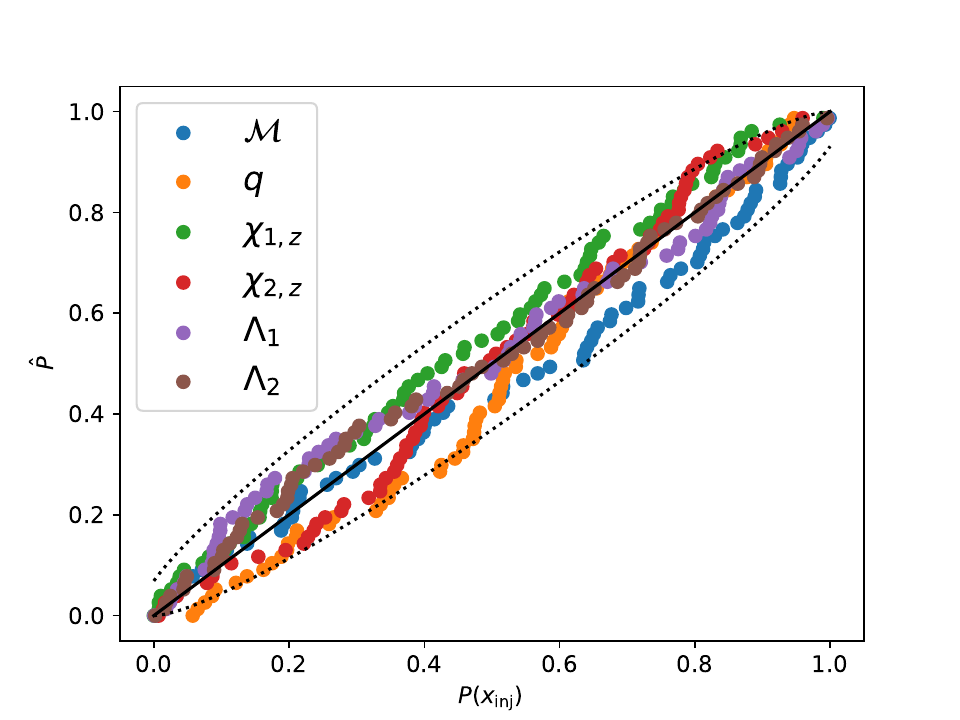}
\includegraphics[scale=0.5]{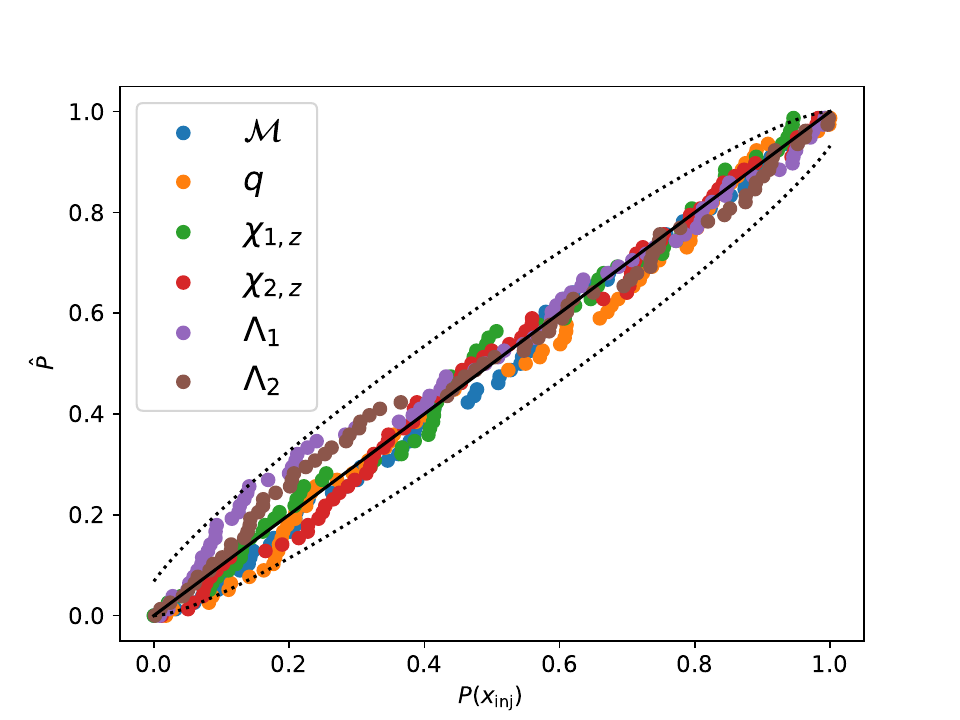}
%\caption{PP-plot of low-amplitude injections with \NRSurTidal with all higher-order-modes($\ell_{max}=5$) and recovered with different waveform models, starting from left, \NRSurTidal $\ell_{max}=5$ and \IMRPD. 
\caption{PP-plot of low-amplitude injections with \NRSurTidal with all higher-order-modes($\ell_{max}=5$)  on the left and \Resum with all higher-order-modes($\ell_{max}=4$) on the right recovered with the same model as the injection. Both sets have the same injection parameters.
The dashed line indicates the 90\% credible interval expected for a cumulative distribution drawn from 100
uniformly-distributed samples.
}
\label{pp-samemodel}
\end{figure*}

\begin{figure*}
\includegraphics[scale=0.5]{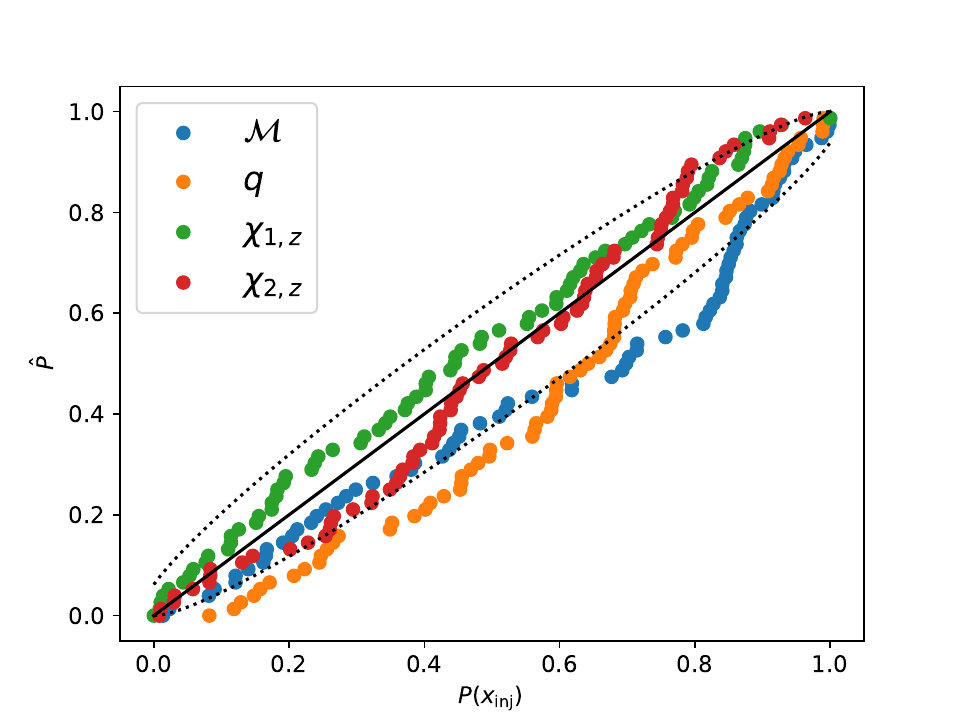}
\includegraphics[scale=0.5]{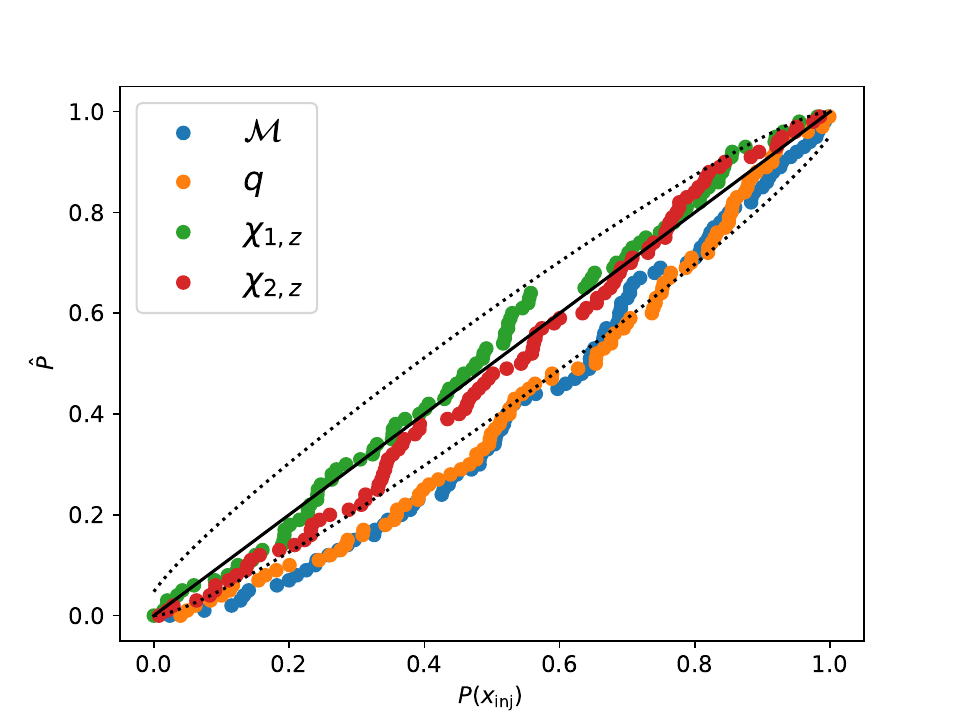}
%\caption{PP-plot of low-amplitude injections with \Resum with all higher-order-modes($\ell_{max}=4$) and recovered with the same waveform model and higher-order-modes setting. 
\caption{PP-plot of low-amplitude injections with \NRSurTidal with all higher-order-modes($\ell_{max}=5$) on the left and \Resum with all higher-order-modes($\ell_{max}=4$) on the right recovered with \IMRPD. Both sets have the same injection parameters.
The dashed line indicates the 90\% credible interval expected for a cumulative distribution drawn from 100
uniformly-distributed samples.
}
\label{pp-IMRPDrec}
\end{figure*}

Figures ~\ref{pp-samemodel} and ~\ref{pp-IMRPDrec} provide another representation of the analyses presented above in the context of JS
divergence: synthetic sources generated with the  \NRSurTidal $\ell_{max}=5$ and \Resum($\ell_{max}=4$) models. In each panel, colored dots show the empirical cumulative distribution of the posterior quantiles of the injections -- the PP plot for each parameter, with colors corresponding to the parameters indicated in the legend. Fig. ~\ref{pp-samemodel} in which the same model was used for both injection
and recovery for a particular panel, we see PP plots for every parameter are consistent with  $P(<p)=p$, as expected. However, Fig. ~\ref{pp-IMRPDrec} where analyses used \IMRPD for recovery, shows that omitting tidal physics entirely can bring in distinct inconsistencies with $P=p$ for injections where tides are significant and important. 

%This emphasizes the choice to

\section{Conclusions}
\label{sec:conclude}

In this paper, we demonstrated that the interpretation of typical low-amplitude BNS sources will frequently exhibit noteworthy
differences, depending on the adopted model for analysis. Specifically, we showed that a JS divergence between
inferences constructed between two different state-of-the-art waveforms would be larger than $10^{-1}$ for a significant population of mergers.
These large differences persist even though the mismatch between the dominant ($2,\pm 2$) mode of this state-of-the-art
waveforms are small, and even though the SNR of our test sources is low.
Additionally, corroborating previous work ~\cite{Dudi:2018jzn,Narikawa:2023deu}, we demonstrate that tidal effects are
essential to include even in interpreting our
population of modest-SNR sources.  Specifically, we showed that neglecting tidal physics in parameter inference causes a PP plot to
deviate significantly from the expected diagonal behavior, indicating a biased recovery of mass and/or spin parameters. 

Our study stands in contrast with the expectations of several previous studies, which have argued that
the effects of waveform systematics for these low-mass, low-amplitude sources will be small.   For example, investigations done in ~\cite{Gamba:2020wgg} suggest that systematic differences will supersede statistical differences
for sources only for high SNR for the current GW detectors. 
%\editremark{XXXX} performed a parameter inference study and found \editremark{ZZZ}. 

Our investigation only demonstrated notable differences in the conclusions derived from different waveform models.
Further study is required to assess to what extent these differences propagate into conclusions derived from a
population of sources or if they average out over the population.

\begin{acknowledgements}
% ROS-  PHY-1707965

The authors would like to thank Carl-Johan Haster for helpful comments.
ROS  gratefully acknowledges support from NSF awards NSF PHY-1912632, PHY-2012057, and AST-1909534.
AY acknowledges support from NSF PHY-2012057 grant. This material is based upon work supported by NSF’s LIGO Laboratory 
    which is a major facility fully funded by the
    National Science Foundation.
This research has made use of data,
    software and/or web tools obtained from the Gravitational Wave 
    Open Science Center (https://www.gw-openscience.org/ ),
    a service of LIGO Laboratory,
    the LIGO Scientific Collaboration and the Virgo Collaboration.
LIGO Laboratory and Advanced LIGO are funded by the 
    United States National Science Foundation (NSF) as
    well as the Science and Technology Facilities Council (STFC) 
    of the United Kingdom,
    the Max-Planck-Society (MPS), 
    and the State of Niedersachsen/Germany 
    for support of the construction of Advanced LIGO 
    and construction and operation of the GEO600 detector.
Additional support for Advanced LIGO was provided by the 
    Australian Research Council.
Virgo is funded through the European Gravitational Observatory (EGO),
    by the French Centre National de Recherche Scientifique (CNRS),
    the Italian Istituto Nazionale di Fisica Nucleare (INFN),
    and the Dutch Nikhef,
    with contributions by institutions from Belgium, Germany, Greece, Hungary,
    Ireland, Japan, Monaco, Poland, Portugal, Spain.
The authors are grateful for computational resources provided by the 
    LIGO Laboratory and supported by National Science Foundation Grants
    PHY-0757058, PHY-0823459 and PHY-1626190, and IUCAA LDG cluster Sarathi.

%The authors are grateful for computational resources provided by 
%the LIGO Laboratories at CIT and LHO supported by National Science
%Foundation Grants PHY0757058 and PHY-0823459.

%\editremark{get boilerplate text}
\end{acknowledgements}

\bibliography{references}
\end{document}